# 1-MHz operation of 1.7-cycle multiple plate compression at 35-W average output power

TAKUYA OKAMOTO,[1,*] YOJI KUNIHASHI,[1] YASUSHI SHINOHARA,[1] HARUKI SANADA,[1] MING-CHANG CHEN,[2] AND KATSUYA OGURI[1]

[1]*NTT Basic Research Laboratories, NTT Corporation, 3-1 Morinosato Wakamiya, Atsugi, Kanagawa 243-0198, Japan*
[2]*Institute of Photonics Technologies, National Tsing Hua University, Hsinchu, Taiwan*
*Corresponding author: tky.okamoto@ntt.com*

**We generated 1.7-cycle and 35-μJ pulses at a 1-MHz repetition rate by using two-stage multiple plate continuum compression of Yb-laser pulses with 80-W average input power. By adjusting the plate positions with careful consideration of the thermal lensing effect due to the high average power, we compressed the output pulse with a 184-fs initial duration to 5.7 fs by using only group-delay-dispersion compensation. This pulse achieved a sufficient beam quality ($M^2 < 1.5$) reaching a focused intensity over $10^{14}$ W/cm$^2$ and a high spatial-spectral homogeneity (98%). Our study holds promise for a MHz-isolated-attosecond-pulse source for advanced attosecond spectroscopic and imaging technologies with unprecedentedly high signal-to-noise ratios.**

Intense few-cycle pulse generation has advanced the frontier of isolated attosecond pulse (IAP) generation technology in the last decade [1–4]. In order to apply IAP to attosecond time-resolved spectroscopy, one of the most important factors is the average photon flux of the source. The photon flux is closely related to the quality of the measurement in terms of the signal-to-noise ratios, detection sensitivity, and measurement time. Scaling up the repetition rate and average power of few-cycle pulse sources is the most direct way of increasing the IAP photon flux [5,6]. Ti:sapphire (Ti:Sa) lasers in combination with various post-compression techniques have been used as few-cycle pulse sources [7–9], which typically have repetition rates of a few kHz and outputs of less than 10 W. If one could make the repetition rate and average power higher, it would overcome the limitations of current IAP sources and lead to novel attosecond spectroscopic applications.

Yb-based lasers are expected to be alternatives to Ti:Sa lasers for scaling up few-cycle pulse generation. Compared with Ti:Sa lasers, they can generate much higher average power, from several tens to hundreds of W, enabling sub-mJ pulses to be obtained at MHz-order repetition rates. However, it is still a challenging task to compress a Yb-based laser pulse into the few-cycle regime, because the gain bandwidth of the Yb-doped medium limits the intrinsic pulse duration to 0.1 – 1 ps [10,11], which is much longer than the typical duration of 20–30 fs for Ti:Sa lasers. To shorten the longer pulse duration of Yb-based lasers, several post-compression schemes, such as using a hollow-core fiber (HCF) [12,13], a multi-pass cell [14–17], or a multiple plate continuum (MPC) [9,18–20], have been investigated [21].

These schemes are classified into two categories: the multi-pass and the single-pass configurations. The multi-pass configuration has been considered challenging for generating sub-two-cycle (or shorter) pulse because it relies on advanced coating technologies of dielectric mirrors with broadband high reflectivity, precise group-delay-dispersion (GDD) management, and high damage threshold despite its excellent throughput of nearly 80% or more [14,15]. To overcome this limitation, the combination of broadband-dielectric and metal-coated mirrors are recently used in the cascaded cell [15,17]. Although this technique achieved the few-cycle compression (6.9 fs), but it required the special metal-coated mirror array on the substrate featuring a low heat expansion and a high heat conductivity, and the water-cooling for 100-W-level average power operation [15]. In contrast, the single-pass configuration is not limited by the bandwidth of optics for generating sub-two-cycle pulses. The HCF scheme is one of the promising candidates [12, 13]. However, it also requires cooling to reduce the thermal effect on the complete straightness of HCF for 100-W-level average power operation [12,13]. The other scheme based on the kagome photonic crystal HCF achieved 1.3-cycle pulses at 10 MHz [22], but the compressed pulse energy of a few μJ was not sufficient for IAP generation.

The MPC scheme, which is categorized in the single-pass scheme, is an alternative way to generate intense few-cycle pulses using high-power Yb-based lasers. This is because the MPC scheme has no limitation of the generating bandwidth and no need of the special treatment for reducing thermal problem such as a mirror substrate featuring low heat expansion and high heat conductivity, and a mirror cooling, which is considered to be drawbacks of the multi-pass-cell compression scheme. In addition, it has potential to achieve higher throughput, greater robustness with a smaller size, and easier alignment than the HCF scheme. A few-cycle pulse compression using a two-stage MPC has recently been demonstrated with a 20-W Yb-based laser at a repetition rate of a few kHz [18, 23, 24]. Remarkably, a monocycle-pulse capability has been demonstrated through precise phase compensation with a 4-f spatial-light-modulator pulse shaper at a 1-kHz repetition rate [19]. Since high-power Yb laser technology is rapidly growing [10,11], it is desired to expand the MPC scheme for few-cycle-pulse compression with average powers up to the 100-W level.

It remains one elusive problem to extend the MPC scheme to high-power pulse compression at high repetition rates – how to handle thermal effects. Because of heat accumulating at the plate [25], it is unclear how much power the MPC method can sustain. In particular, the long pulse duration of Yb-based lasers compared with that of Ti:Sa lasers is more likely to induce such undesired effects. Moreover, the resulting thermal lens may influence both wavefront and propagation of the beam, altering the following nonlinear processes, e.g., self-phase modulation and self-steepening [9, 19], in the MPC. To achieve few cycle-pulse compression with high average power, it is therefore important to suppress these excessive nonlinear effects that cause the complex higher-order dispersion in the spectral broadening process.

Here, we demonstrate that the MPC scheme can generate 1.7-cycle pulses at a 1-MHz repetition rate and 80-W input average power. The scheme reported here involves an elaborate suppression of higher-order dispersion that takes into account the thermal lensing effect induced by high average powers. The 1.7-cycle pulses provide a focused peak intensity exceeding $10^{14}$ W/cm$^2$, promising MHz attosecond pulse generation.

We employed a two-stage MPC compressor consisting of fused silica plates with various thicknesses and a chirped mirrors with a negative GDD, as shown in Fig. 1. The laser was a 1035-nm Yb:KGW amplifier (CARBIDE, Light Conversion, Ltd.) with a maximum average power of 80 W. It can operate from single shot to 2 MHz with variable pulse energy up to 1 mJ. For the experiment on few-cycle-pulse compression, we set an initial pulse peak power of 0.43 GW (80 µJ over a pulse duration of 184 fs) at a 1-MHz repetition rate.

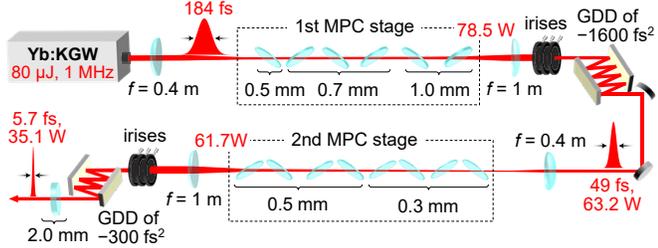

Fig. 1. Schematic diagram of the experimental setup. $f$, focal length.

The first MPC stage employed six fused silica plates with individual thicknesses of 0.5, 0.7, or 1.0 mm set at the Brewster angle. By placing a self-focusing beam in the free space between the plates, optical breakdown was avoided while ensuring strong nonlinearity in each plate [9]. The plates were gradually increased in thickness to efficiently accumulate a nonlinear phase shift, i.e., the B integral. The initial beam was focused by a dielectric concave mirror with a focal length of 40 cm. The B integral was about 1.0 rad at the first plate placed 1.7 cm after the focus, which was estimated by a spot-size measurement. The spacings between the following plates were 1.2, 1.8, 2.1, 2.3, and 3.0 cm. The footprint of the first MPC stage was only about 3.8 × 11 cm$^2$, which is significantly more compact than that of the HCF and multi-pass cell systems for high-average-power operation [12, 13, 15].

The initial pulses were compressed by the first MPC stage. The spectra transmitted through different numbers of plates in the first MPC stage were measured by InGaAs-based spectrometer (waveScan, APE GmbH). Figure 2(a) shows the dependence of the spectra on the number of plates. The spectra clearly show symmetric broadening with a decrease in the fundamental wavelength. This tendency indicates that the spectral broadening mainly originates from self-phase modulation accompanied with a GDD. The resultant 10$^{-3}$ bandwidth ranges approximately from 970 to 1090 nm. The spectrally broadened pulse was collimated by a dielectric concave mirror and then spatially filtered using several irises to remove outer-ring components due to the high-order wavefront distortion induced in the MPC [9,18–20]. The following dispersion compensation was performed with a GDD of −1600 fs$^2$. Figure 2(b) plots the fringe-resolved autocorrelation trace for the compressed pulse. The full width at half maximum (FWHM) of the autocorrelation width was 83 fs, corresponding to 49 fs for a sech$^2$ pulse shape. The beam quality was evaluated by measuring $M^2$ in accordance with the D4$\sigma$ definition in the ISO 11670 standard (Fig. 2(c)). The focusing component was a plano-convex lens with a focal length of 50 cm, and the beam profiler was a Si-CCD camera with a sensitivity from 190 to 1100 nm. The horizontal ($M^2_x$) and vertical ($M^2_y$) beam quality factors were measured to be 1.09 and 1.06, respectively, which are almost unchanged from the input beam quality ($M^2 = 1.08 \times 1.08$). The average power of the pulse was measured over the course of 3 hours to be 63.2 ± 0.4 W (Fig. 2(d)), yielding a conversion efficiency of 79.0%.

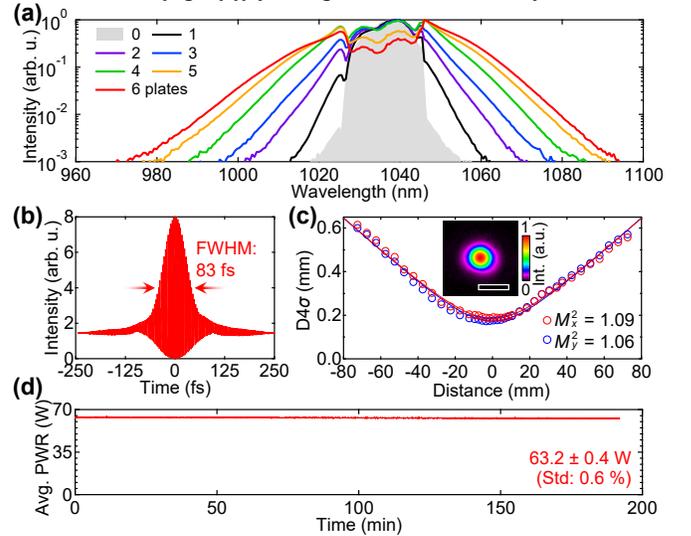

Fig. 2. Characterization of pulses compressed by the first MPC compressor. (a) Output spectra for different numbers of plates. (b) Fringe-resolved autocorrelation trace. (c) $M^2$ measurement in accordance with the D4$\sigma$ definition. Inset: beam profile at a focus. The scale bar corresponds to 200 µm. (d) Output power measured over 3 hours.

To investigate the thermal effects, we examined changes in the compressed pulse as the repetition rate varied while keeping the pulse energy constant. Figure 3(a) displays a representative change in the autocorrelation traces at repetition rates of 1 and 0.1 MHz, corresponding to average powers of 80 and 8 W, respectively. An apparent increase in the detrimental pedestal around the main lobe is visible as the repetition rate decreases. The pedestal monotonically increases from 12.0 to 26.8% as the repetition rate decreases from 1 to 0.1 MHz (inset of Fig. 3(a)). Since regular chirped mirrors with a constant GDD were unable to eliminate the pedestal, it is likely due to higher-order dispersion caused by excessive nonlinear effects such as self-phase modulation or self-steepening. This temporal modification can be explained by the deviation of the beam waist position optimized for 80 W, as shown in Fig. 3(b). As the repetition rate decreases, the thermal lensing effect should weaken, leading to higher intensity in the following plate and causing higher-order dispersion. We observed a ~

3-mm forward shift of the beam waist after the first plate as the repetition rate decreased from 1 to 0.1 MHz. Importantly, the optimum pulse condition, i.e., the same duration and pedestal at 1 MHz, can be recovered by adjusting the MPC stages. It indicates that the thermal effect can be easily compensated by optimizing the plate positions, even if changing the input average power causes the shift of the focal point. These results point out that, including thermal effects, the fused-silica-based MPC scheme can be scaled up to sub-100 W at MHz-repetition rates.

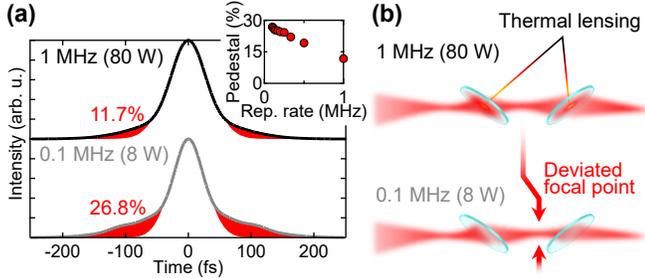

Fig. 3. Thermal effects in the MPC stage. The plate positions were optimized for 1 MHz. (a) Comparison of autocorrelation traces at 0.1 and 1 MHz. The red regions indicate the pedestals. Inset: pedestal percentage versus repetition rate. (b) Illustration of deviation of beam waist due to thermal lensing effect.

To reach the few-cycle regime, the pulse from the first MPC stage was compressed again with the second MPC stage. First, it was focused into the second MPC stage by a plano-convex lens with a focal length of 40 cm. We made use of a similar plate arrangement as in the first MPC stage, placing four 0.3-mm plates and four thicker 0.5-mm plates sequentially. The first plate was set 20 mm away from the focus, yielding a B integral of ~1.2 rad. The optimized spacings between the plates were 13, 13, 12, 12, 12, 17, and 15 mm. Figure 4(a) plots the dependence of the spectra on the number of plates. The spectrum after the last (8th) plate was slightly asymmetric because of the self-steepening effect [9, 19]. Adding more plates did not cause any further spectral broadening, but only distorted the spectra. The final spectrum exhibited octave spanning at the $10^{-3}$ level, reaching the visible region, as shown in the insets in Fig. 4(a). After the broadening, the outer-ring component of the beam was spatially filtered as depicted in the lower inset in Fig. 4(a). The final compression was performed at the sixth bounce on the broadband chirped mirror pairs with a GDD of −50 $fs^2$ and passage through 2-mm fused silica with a GDD of 37 $fs^2$ at a wavelength of 1035 nm. The final compressed pulses were characterized using second-harmonic frequency resolved optical gating (SHG-FROG). The measured and retrieved SHG-FROG traces are presented in Fig. 4(b). The spectral phase, plotted in Fig. 4(c), exhibited good GDD compensation (only 12 $fs^2$) and slight residual high-order dispersion. The FWHM of the temporal pulse envelope was 5.7 fs (Fig. 4(d)), which was within 10% of the Fourier-transform-limited (FTL) pulse duration. This duration corresponds to a 1.7 optical cycle at a carrier wavelength of 1035 nm. The main lobe of the pulse envelope (depicted as the filled area in Fig. 4(d)) was 63.0 ± 2.2%. Consequently, the final peak power of 3.9 GW was achieved, thus resulting in the peak power boost of 9.0, defined as the ratio of output to input peak power. Note that this sub-two-cycle pulse compression was achieved only by GDD compensation, which contrasts with the conventional way in which special attention must be paid to the large high-order nonlinearity by the MPC compression [19]. This result indicates that carefully choosing the thicknesses, positions, and number of the fused silica plates enabled us to induce an ideal self-phase modulation, leading to the symmetric spectral broadening shown in Fig. 4 (a). The available average power was measured to be 35.1 ± 0.4 W in a 1-hour measurement (Fig. 4(e)). The throughput of the second MPC compressor was 55.6%, yielding a total throughput of 43.9%. In comparison with the previous results of MHz-scale two-cycle-level pulse compression, we achieved better performance in pulse duration and total throughput (conversion efficiency) than the HCF scheme (6.3 fs and 33% at 1.27 MHz [12]) and the optical parametric chirped-pulse amplification scheme (6.6 fs (wavelength of 918 nm) and 15% at 0.6 MHz for [5]), but could not surpass the total throughput of the multi-pass-cell scheme (6.9 fs and 78% at 0.5 MHz [15]).

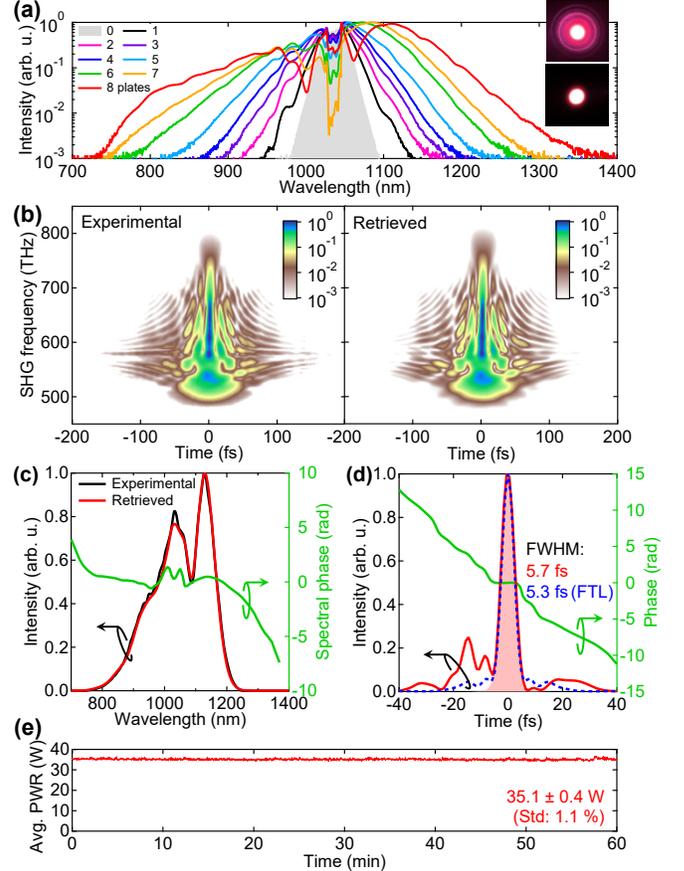

Fig. 4. Characterization of pulses compressed at the second MPC compressor. (a) Spectral broadening with different numbers of plates. Insets: photographic images of spectrally broadened beam without (upper) and with (lower) spatial filters. (b) Experimental and retrieved SHG-FROG traces. The FROG error was 2.0% over a 512×512 retrieval grid. (c) Retrieved (red), experimental (black) intensity spectrum and spectral phase (green). (d) Retrieved temporal pulse intensity (red), phase (green), and FTL pulse intensity (dashed blue). The filled region indicates the main lobe. (e) Average power measured in 1 hour.

Moreover, we measured the spatial-spectral homogeneity to quantify the 1.7-cycle pulses. The spatial-spectral homogeneity map was sampled along the horizontal direction with a 150-μm-diameter pinhole, as shown in Fig. 5(a). The resultant integrated power and homogeneity were calculated (Fig. 5(b)), as defined in [26]. The measurement confirmed the high homogeneity of the first (97.5%) and second (97.9%) MPC stages, which maintains almost the same value as the input beam (97.7%). These results demonstrate no significant degradation in the spatial-spectral homogeneity by the MPC compression at the 80-W operation.

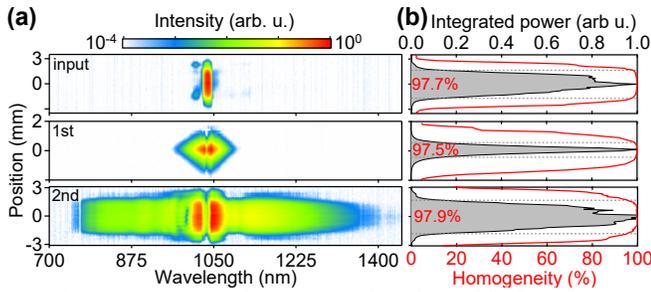

Fig. 5. Homogeneity of compressed pulses. (a) Spatial-spectral distribution of the pulses along the horizontal direction. (b) Spatial-spectral beam homogeneity and integrated power. Percentages are effective homogeneities within the $1/e^2$ level (dashed lines).

Finally, we checked the focusing quality of the 1.7-cycle pulses for attosecond pulse generation. Figure 6(a) displays the $M^2$ measurement, indicating an $M^2$ of 1.46 × 1.48, assuming the central wavelength of 1035 nm. While considering the quantum efficiency curve of the Si-CCD, the $M^2$ value fitting becomes 1.61 × 1.64 since the relative-response-weighted wavelength shifts to 931 nm. Despite the degradation in the $M^2$, the focus quality looks excellent. We verified the applicability of the 1.7-cycle pulse to attosecond pulse generation by tightly focusing it with an off-axis parabolic mirror with a focal length of 10 cm, as shown in Fig 6(b). The $1/e^2$-diameter area was 27.7 × 25.3 μm$^2$ and had a small pedestal of 7.8%. The estimated focused peak intensity for the main lobe (Fig. 4(d)) is $7.1 \times 10^{14}$ W/cm$^2$ with an assumption of the same focusing quality for a longer wavelength than 1100 nm. This focused pulse can easily ionize air as depicted in Fig 6(c), revealing that it is well beyond the typical value of $10^{14}$ W/cm$^2$ required for attosecond pulse generation.

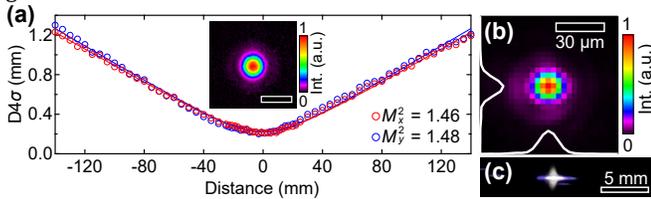

Fig. 6. Focusing quality of the 1.7-cycle pulse. (a) $M^2$ measurement. The inset displays the beam profile at a focus. The scale bar indicates 200 μm. (b) Tightly focused beam profile with line profiles (white curves). (c) Photographic image of the air breakdown.

In summary, we generated the 1.7-cycle pulses with an energy of 35.1 μJ and a 1-MHz repetition rate by using the two-stage MPC compressor. This achievement clearly confirmed that the MPC compression is a simple, compact, and low-cost scheme for realizing sub-two-cycle regime while supporting sub-100-W average power at MHz-order operation. The 1.7-cycle pulses provide a sufficient beam quality reaching a focused intensity in excess of $10^{14}$ W/cm$^2$ and a high spatial-spectral homogeneity. Thus, this development is a milestone in MHz-IAP generation and may lead to attosecond spectroscopic and imaging technologies with unprecedentedly high signal-to-noise ratios and sensitivities.

**Funding.** JSPS KAKENHI Grant Numbers JP20H02563 and JP20H05670. Taiwan NSTC Grant Number 111-2628-M-007-001.

**Disclosures.** The authors declare no conflicts of interest.

**Data availability.** Data underlying the presented results may be obtained from the corresponding author upon reasonable request.